\documentclass[]{pasj01}

\begin{document} 
\Received{}
\Accepted{}

\title{Suzaku and Chandra observations of the galaxy cluster RXC J1053.7+5453 with a radio relic}

\author{Madoka \textsc{Itahana}\altaffilmark{1}}
\altaffiltext{1}{School of Science and Engineering, Yamagata University, 
                  Kojirakawa-machi 1-4-12, Yamagata 990-8560, Japan}
\email{itahana@ksirius.kj.yamagata-u.ac.jp}

\author{Motokazu \textsc{Takizawa}\altaffilmark{2}}
\altaffiltext{2}{Department of Physics, Yamagata University, Kojirakawa-machi
                  1-4-12, Yamagata 990-8560, Japan}
\email{takizawa@sci.kj.yamagata-u.ac.jp}

\author{Hiroki \textsc{Akamatsu}\altaffilmark{3}}
\altaffiltext{3}{SRON Netherlands Institute for Space Research, Sorbonnelaan 2, 3584 CA Utrecht, The Netherlands}

\author{Reinout J. \textsc{van Weeren}\altaffilmark{4}}
\altaffiltext{4}{Harvard-Smithsonian Center for Astrophysics, 60 Garden Street, Cambridge, MA 02138, USA}

\author{Hajime \textsc{Kawahara}\altaffilmark{5,6}}
\altaffiltext{5}{Department of Earth and Planetary Science, The University of Tokyo, 7-3-1 Hongo, Bunkyo-ku, Tokyo 133-0033, Japan}
\altaffiltext{6}{Research Center for the Early Universe, School of Science, The University of Tokyo, Tokyo 113-0033, Japan}

\author{Yasushi \textsc{Fukazawa}\altaffilmark{7}}
\altaffiltext{7}{ Hiroshima University, 1-3-1 Kagamiyama, Higashi-Hiroshima, Hiroshima 739-8526, Japan}

\author{Jelle S. \textsc{Kaastra}\altaffilmark{3,11}}

\author{Kazuhiro \textsc{Nakazawa}\altaffilmark{8}}
\altaffiltext{8}{Department of Physics, The University of Tokyo, 7-3-1 Hongo, Bunkyo-ku, Tokyo 133-0033, Japan}

\author{Takaya \textsc{Ohashi}\altaffilmark{9}}
\altaffiltext{9}{Department of Physics, Tokyo Metropolitan University, 1-1 Minami-Osawa, Hachioji, Tokyo 192-0397, Japan}

\author{Naomi \textsc{Ota}\altaffilmark{10}}
\altaffiltext{10}{Department of Physics, Nara Women's University, Kitauoyanishi-machi, Nara 630-8506, Japan}

\author{Huub J. A. \textsc{R\"{o}ttgering}\altaffilmark{11}}
\altaffiltext{11}{Leiden Observatory, Leiden University, PO Box 9513, 2300 RA Leiden, The Netherlands}

\author{Jacco \textsc{Vink}\altaffilmark{12}}
\altaffiltext{12}{Anton Pannekoek Institute\/GRAPPA, University of Amsterdam, PO Box 94249, NL-1090 GE Amsterdam, the Netherlands}

\author{Fabio \textsc{Zandanel}\altaffilmark{13}}
\altaffiltext{13}{GRAPPA Institute, University of Amsterdam, 1098 XH Amsterd am, The Netherlands}



\maketitle

\begin{abstract}
We present the results of Suzaku and Chandra observations of the galaxy cluster RXC J1053.7+5453 ($z=0.0704$), which contains a radio relic. 
The radio relic is located at the distance of $\sim 540$ kpc from the X-ray peak toward the west.
We measured the temperature of this cluster for the first time.
The resultant temperature in the center is $ \sim 1.3$ keV, which is lower than the value expected from the X-ray luminosity - temperature and the velocity dispersion - temperature relation.
Though we did not find a significant temperature jump at the outer edge of the relic, our results suggest that the temperature decreases outward across the relic.
Assuming the existence of the shock at the relic, its Mach number becomes $M \simeq 1.4 $.
A possible spatial variation of Mach number along the relic is suggested.
Additionally, a sharp surface brightness edge is found at the distance of $\sim 160$ kpc from the X-ray peak toward the west in the Chandra image. 
We performed X-ray spectral and surface brightness analyses around the edge with Suzaku and Chandra data, respectively.
The obtained surface brightness and temperature profiles suggest that this edge is not a shock but likely a cold front.  
Alternatively, it cannot be ruled out that thermal pressure is really discontinuous across the edge.
In this case, if the pressure across the surface brightness edge is in equilibrium, other forms of pressure sources, such as cosmic-rays, are necessary. 
We searched for the non-thermal inverse Compton component in the relic region.
Assuming the photon index $ \Gamma = 2.0$, the resultant upper limit of the flux is $1.9 \times 10^{-14} {\rm erg \ s^{-1} \ cm^{-2}}$ for $4.50 \times 10^{-3} {\rm \ deg^{2}}$ area in the 0.3-10 keV band, which implies that the lower limit of magnetic field strength becomes $ 0.7 {\rm \ \mu G}$.

\end{abstract}

\section{Introduction}
According to the standard theory of structure formation in the universe, galaxy clusters are built up through mergers and absorption of smaller galaxy clusters and groups.
Cluster major mergers are the most energetic phenomena ($ \sim 10^{64}$ erg) in the universe.
Edge-like features are found in X-ray images obtained by Chandra and XMM-Newton for some clusters.
Such structures are believed to be a shock or cold front.
Temperature, density and pressure profiles show a discontinuity across a shock front.
Similarly, the temperature and density are discontinuous, but the pressure is continuous across a cold front.
It is believed that these are the evidences of a cluster merger.
Numerical simulations show that shocks and contact discontinuities appear in the intracluster medium (ICM) during cluster major mergers \citep{Ricker01,Takizawa05,Akahori10,Takizawa10}.
The features of a contact discontinuity agree with those of cold fronts.
While a large part of the kinetic energy is converted into the thermal ICM during cluster mergers, some part of it will be converted into a non-thermal form such as magnetic fields and cosmic-rays \citep{Ohno02,Takizawa08,Zuhone11,Donnert13}.
Some merging clusters host diffuse non-thermal radio sources \citep{Feretti12,Brunetti14}.
This radio emission is from synchrotron radiation by the interaction of relativistic electrons whose energy is $\sim {\rm GeV}$, and magnetic field of $\sim {\rm \mu G}$ in the ICM, which are classified into three categories by their size, morphology and location. 
Radio halos are in the central part of the cluster and have  similar morphology to the ICM X-ray emission \citep{Feretti97,Govoni04,vanWeeren11,Scaife15}.
Mini halos are also located in the cluster center, but their typical size is much smaller than halos \citep{Feretti12,Giacintucci14}.
On the other hand, radio relics are usually in cluster outskirts and show an arc-like shape \citep{Rottgering97,Bonafede09,vanWeeren10,vanWeeren16}.

Owing to their locations and morphology, 
it is believed that radio relics should have a close connection with shock fronts caused by cluster mergers.
Thus, radio relics can be good tracers of merger shocks in clusters and the shocks are expected to be located at the outer edge of the relics.
Recently, in fact, temperature and density jumps are found across the relic outer edge in some clusters by the X-ray observations \citep{Finoguenov10,Akamatsu12a,Akamatsu13,Ogrean13,Eckert16,Akamatsu16}.
This is direct evidence of the relationship between relics and shocks.
Mach numbers of shocks can be estimated from radio and X-ray observations independently.  
From radio observations, we can obtain radio spectral index at the outer edge of relics.
Assuming a simple diffusive shock acceleration (DSA) theory \citep{Drury83,Blandford87}, this spectral index can be related with the shock Mach numbers \citep{Rybicki79} .
From X-ray observations, we can obtain the temperature and density distribution around relics.
We can estimate the Mach number with Rankine-Hugoniot conditions from the temperature and density jump at shock front \citep{Landau59,Shu92}.
If the simple DSA model is true, consistent results will be obtained through both methods.

The high energy electrons in synchrotron radiation regions (radio relics and halos) emit non-thermal X-rays via the inverse Compton scattering of cosmic microwave background (CMB) photons \citep{Bartels15}. 
By comparing  synchrotron radio and non-thermal X-ray fluxes , we are able to estimate the magnetic field strength in the non-thermal radio emission regions.
Even if we obtained only the upper limit of non-thermal X-ray flux, we can estimate the lower limit of the magnetic field strength.
Though a lot of attempts to search for the non-thermal X-ray components have been done, no firm detections are reported \citep{Ajello09,Nakazawa09,Sugawara09,Itahana15,Akamatsu16}.

The galaxy cluster RXC J1053.7+5453 ($z=0.0704$) is known to host a radio relic.
The radio relic is located at the distance of $542$ kpc from the X-ray peak toward the west.
Its X-ray luminosity with ROSAT is $L_{\rm X[0.1 - 2.4 keV]}=0.96 \times 10^{44} {\rm \ erg/s}$ \citep{popesso04}.
\citet{Ebeling98} estimated the temperature $kT \sim 3 {\rm \ keV}$ from $L_{\rm X} - kT$ relation.
This value is rather low for clusters with radio relics.
However, there is no direct temperature measurement for this cluster.
According to the Sloan Digital Sky Survey (SDSS) data, the velocity dispersion of the member galaxies is $665^{+51}_{-45} {\rm \ km/s}$ and the virial radius is $r_{200} =1.52 {\rm \ Mpc}$ \citep{Aguerri07},
which is defined as the radius within which the mean density becomes 200 times of the critical density of the universe.
From the radio observation, \citet{vanWeeren11} reported that the relic length and the radio flux density are  $600 {\rm \ kpc}$ and $S_{\rm 1382 \ MHz} = 15 \pm 2 {\rm \ mJy}$, respectively.
There is no observational information about the radio spectra of this relic.

\begin{figure}
 \begin{center} 
  \includegraphics[width=8cm]{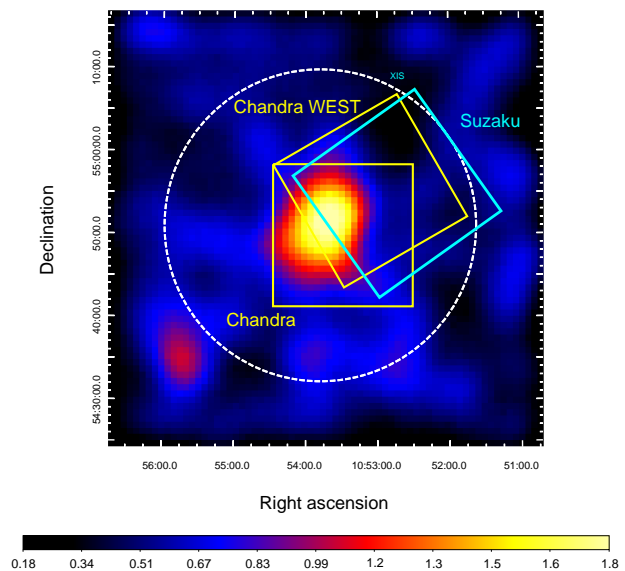} 
 \end{center}
\caption{A ROSAT image of RXC J1053.7+5453 in the 0.1-2.4 keV band. The image intensity is arbitrary unit and rms value of the image fluctuation is $ \sigma = 0.48$. The yellow and light blue boxes show the FOVs of the Chandra ACIS and Suzaku XIS observations. The white dashed circle shows the virial radius of RXC J1053.7+5453 ($r_{200} =18.'8$ ). }\label{fig:0}
\end{figure}

We observed a field around a radio relic in the galaxy cluster RXC J1053.7+5453 with Suzaku.
Suzaku is more suitable for spectral analysis of the ICM in the outer parts of the cluster, because the mounted X-ray Imaging Spectrometer (XIS) has a good sensitivity for low surface brightness diffuse sources and low and stable background \citep{Mitsuda07}.
Additionally, we used Chandra archive data for the surface brightness analysis and the point sources removal, because the Advanced CCD Imaging Spectrometer (ACIS) has high spatial resolution.
Figure \ref{fig:0} shows a ROSAT image of RXC J1053.7+5453 in the 0.1-2.4 keV band.
The yellow and light blue boxes show the field of views (FOVs) in the Chandra ACIS and Suzaku XIS observations, respectively.
The white dashed circle shows the virial radius of RXC J1053.7+5453 ($r_{200} =18.'8$ ).

In this paper, we present the Suzaku and Chandra X-ray observations of the galaxy cluster RXC J1053.7+5453.
The outline of the paper is organized as follows.
We describe the observation and data reduction in section 2.
Data analysis and our results are presented in section 3.
We discuss the results in section 4.
We summarize the results in section 5.
We used canonical cosmological parameters of $H_0 = 70 {\rm km \ s^{-1} \ Mpc^{-1}}$, $\Omega_0 = 0.27$, and $\Lambda_0 = 0.73$.
At the redshift of the cluster, $1'$ corresponds to $81$ kpc.
The solar abundances are normalized to \citet{Asplund09}.
Unless otherwise stated, the errors correspond to 90\% confidence level.

\section{Observations and Data Reduction}
We observed a field around the radio relic in galaxy cluster RXC J1053.7+5453 with Suzaku on 2014 November 01 - 03.
This observation is one of the Suzaku AO9 Key projects.
In order to estimate the background components, we used the Suzaku archive data of the Lockman Hole.
We used the data observed on 2009 June 12 - 14 (ID:104002010) because this is the data of the field nearest to this cluster on the sky plain among Lockman Hole observations of Suzaku.
Table \ref{tab:1} shows an observational log of RXC J1053.7+5453 and the Lockman Hole.
RXJ1053 and Lockman Hole data were processed with Suzaku pipeline processing, version 2.8.20.37 and 2.4.12.26, respectively.
We used HEAsoft version 6.19 for the Suzaku data.
The 20150312 calibration data files were adopted.

\begin{figure}
 \begin{center} 
  \includegraphics[width=8cm]{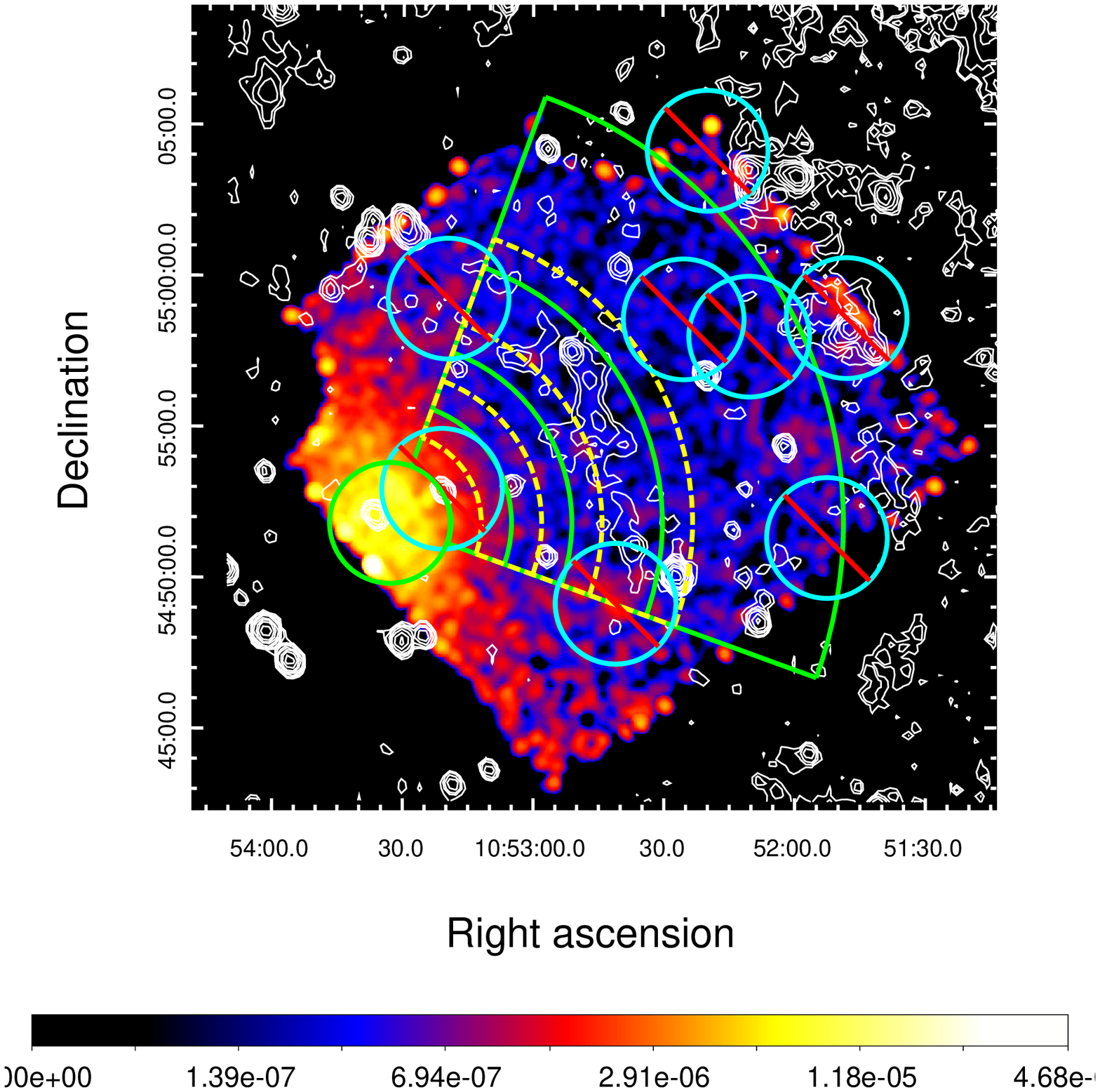} 
  \includegraphics[width=9cm]{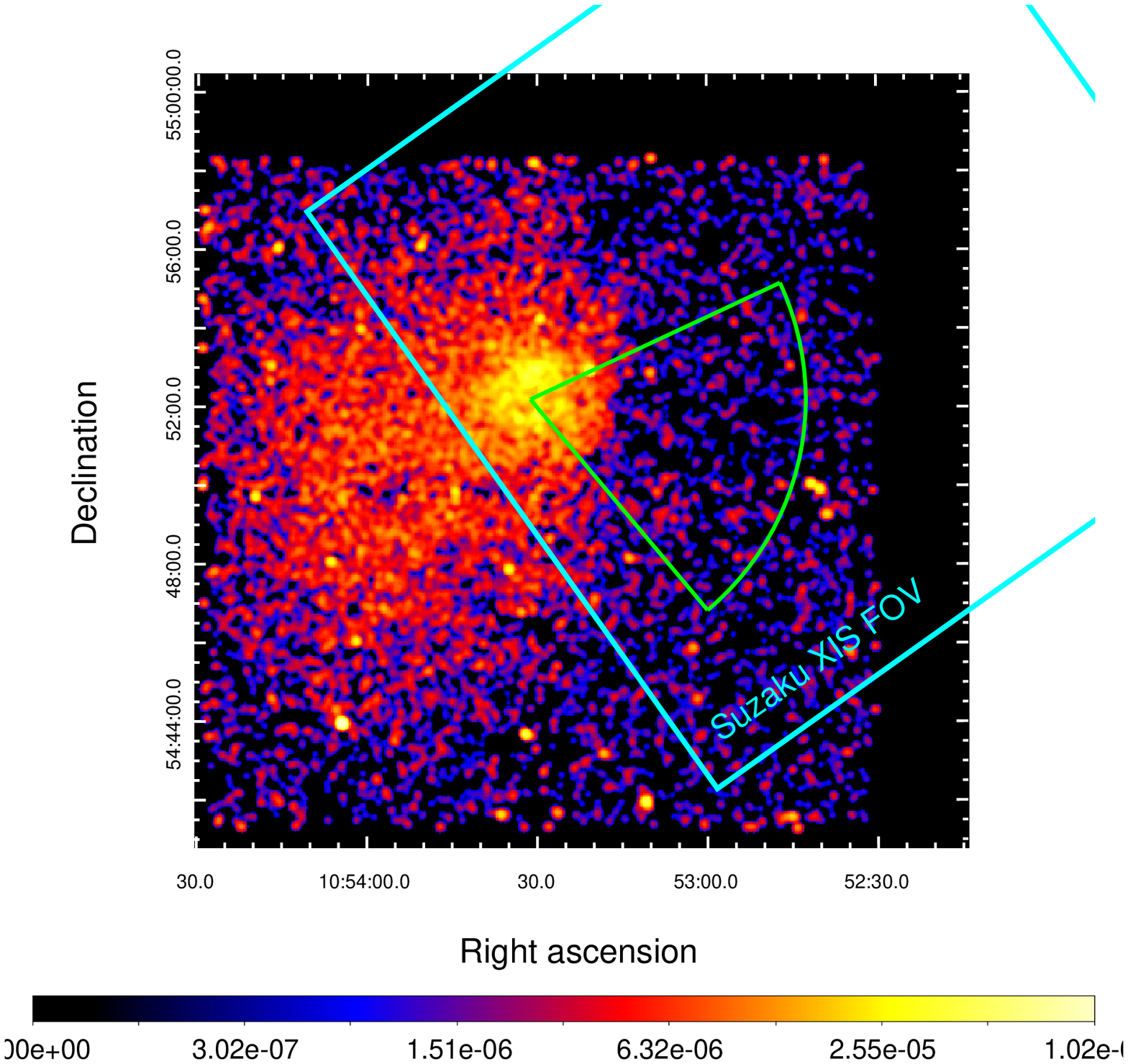} 
 \end{center}
\caption{Top: An XIS1 image in the 0.5-8.0 keV band (Obs.ID:809120010) with the 1382 MHz radio contours \citep{vanWeeren11}. The X-ray image was corrected for exposure and vignetting effects after subtracting NXB and smoothed by a Gaussian kernel with $\sigma=0.'26$. The radio contours are drawn at $ [1,2,4,8,...] \times 0.1 {\rm \ mJy/beam}$. The green and yellow regions were used for the spectral analysis, with annular radius of 2', 4', 6', 9', 15' and 3', 5', 7', 10', respectively. The light blue circles are excluded regions of a point source. Bottom: A Chandra image in 0.5-2.0 keV band (Obs.ID:15322). The image was corrected for exposure. The green sector is a region used for extracting surface brightness profile (in subsection \ref{sec:CF}). Light blue box is the Suzaku XIS FOV.

 }\label{fig:1}
\end{figure}

The XIS data were processed through default screening criteria.
In addition, data obtained in the periods with geomagnetic comic-ray cut-off rigidity ($COR2$)$ > 8.0$ GV were excluded.
As a results, the effective exposure times became 71.6 ks and 63.8 ks for the cluster and Lockman Hole regions, respectively.
In order to reduce the non-X-ray background (NXB) level, which increased after changing the amount of charge injection, we applied additional processing for XIS1 following the processes descriptions in the Suzaku XIS analysis topics
\footnote{http://www.astro.isas.jaxa.jp/suzaku/analysis/xis/xis1\_ci\_6\_nxb/}.
We did not use the XIS0 segment A which is damaged because of a micrometeorite accident 
\footnote{http://www.astro.isas.ac.jp/suzaku/doc/suzakumemo/suzakumemo-2010-01.pdf}.
NXB spectra and images of XIS were generated using the ftool ``xisnxbgen'' \citep{Tawa08}.
Top panel of figure \ref{fig:1} shows a 0.5-8.0 keV XIS1 image with the 1382 MHz radio contours \citep{vanWeeren11}.
The X-ray image was corrected for exposure and vignetting effects after subtracting NXB, and smoothed by a Gaussian kernel with $\sigma=0.'26$.


RXC J1053.7+5453 was also observed with Chandra on 2013 June 22 (ObsID:15322).
In addition, we observed a field around the radio relic with Chandra on 2016 February 09 (ObsID:17207), to search for point sources.
We used CIAO version 4.8 with the calibration files of CALDB version 4.7.0.
The event files were reprocessed using the task ``chandra\_repro''.
The ``lc\_clean'' algorithm was used to filter the soft proton flares, and the light curves were visually inspected to check for any residual flaring.
As a results, the effective exposure times became 24.5 ks and 6 ks for the central and west field, respectively.
A Chandra image in the 0.5-2.0 keV band is shown in the bottom panel of figure \ref{fig:1}.

\begin{table*}
\caption{The observational log of RXC J1053.7+5453 and Lockman Hole.}
\begin{center}
\begin{tabular}{clccc} \hline
       &Name (Obs.ID)                & (RA, Dec)           & Observation Date & Exposure (ks)\footnotemark[$*$] \\ \hline
Suzaku &RXC J1053.7+5453(809120010)  & (163.1807,54.9140)  & 2014/11/01-03    & 71.6           \\ 
       &LOCKMAN HOLE (104002010)     & (162.9375,57.2667)  & 2009/6/12-14     & 63.8           \\ \hline
Chandra&RXC J1053.7+5453(15322)      & (163.4148,54.8296)  & 2013/6/22        & 24.5          \\ 
       &RXC J1053.7+5453 WEST(17207) & (163.2642,54.9077)  & 2016/2/09        & 6           \\ \hline
   \multicolumn{2}{@{}l@{}}{\hbox to 0pt{\parbox{180mm}{\footnotesize
       \footnotemark[$*$] Effective exposure time after data screening as described in the text.
       \par\noindent
     }\hss}}
\end{tabular}
\label{tab:1}
\end{center}
\end{table*}

\section{Data Analysis and Results}  
We used Suzaku data for the spectral analysis.
For the spectral analysis of the XIS data, we generated redistribution matrix files (RMFs) and ancillary response files (ARFs) with the ftool ``xisrmfgen'' and ``xissimarfgen'' \citep{Ishisaki07}, respectively.
Uniform emission over a circular region with $20'$ radius was used as an input image to generate an ARF.

\subsection{The Background Components}
\label{sub:back}
We estimate the background components using the data of the Lockman Hole field.
We assume that the background components are composed of the Local Hot Bubble (LHB), the hot gas of the Milky Way Halo (MWH) and the Cosmic X-ray background (CXB).
Then, we fit the spectrum of the background field using the following model:
\begin{eqnarray}
  apec_{\rm LHB}+phabs*(apec_{\rm MWH}+powerlaw_{\rm CXB}),
\end{eqnarray}
where $apec_{\rm LHB}$, $apec_{\rm MWH}$ and $powerlaw_{\rm CXB}$ represent the LHB, MWH and CXB, respectively.
We fix the temperature of LHB to 0.08 keV, and the redshift and abundance of both the LHB and MWH to zero and solar, respectively.
We assume $N_{\rm H} = 6.05 \times 10^{19} {\rm cm^{-2}}$ for the Galactic absorption \citep{Willingale13}.
The photon index of the CXB is fixed to $\Gamma=1.4$ \citep{Kushino02}.
For the spectral fitting of the background components, we used energy bands of 0.7 - 7.0 keV (XIS0, 3) and 0.6 - 7.0 keV (XIS1).
However, the energy band of 1.7 - 1.8 keV was ignored, because the response matrix around Si-K edge had residual uncertainties.

\begin{figure}
 \begin{center}
  \includegraphics[width=5.5cm,angle=270]{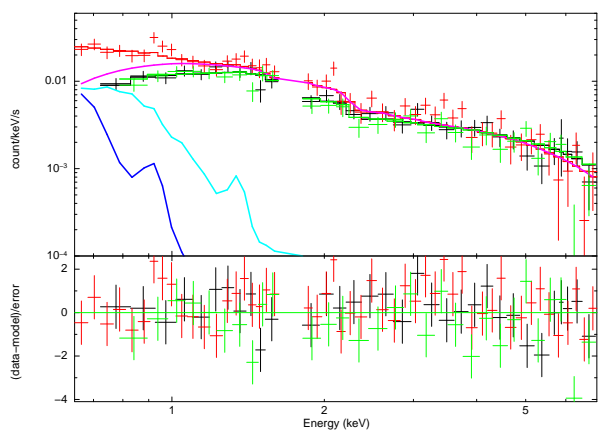} 
 \end{center}
 \caption{The XIS spectra of the background field fitted with the background model described in the text.
 The  black, red, and green crosses show the spectra of XIS0, XIS1, and XIS3, respectively.
The each component and total of the best-fit model spectra are also shown with solid histograms.
 The blue, light blue, and magenta solid histograms represent the LHB, MWH, and CXB components, respectively.}\label{fig:2}
\end{figure}

Figure \ref{fig:2} shows the spectra of the background field fitted with the above-mentioned model, where black, red, and green crosses show the spectra of XIS0, XIS1, and XIS3, respectively.
The each component and total of the best-fit model spectra are also shown with solid histograms.
Table \ref{tab:BGD} shows the detailed results of the best-fit model.
The obtained temperature of the MWH component ($kT = 0.28^{+0.73}_{-0.07}$ keV) is consistent with the typical value ($kT \sim 0.3$ keV; \cite{Yoshino09}).

\begin{table*}
  \caption{Best-fit background parameters for the XIS spectra}\label{tab:BGD}
  \begin{center}
    \begin{tabular}{ccc} \hline 
      Model Component   & Parameter                          & Value
     \\ \hline
      LHB               & $kT$\footnotemark[$*$]             & 0.08 (fixed) \\
                         & $N$\footnotemark[$\dagger$]        & $5.03^{+9.53}_{-5.02} \times 10^{-2}$   \\ \hline
      MWH               & $kT$\footnotemark[$*$]             & $0.28^{+0.73}_{-0.07} $           \\
                        & $N$\footnotemark[$\dagger$]        & $3.98^{+7.94}_{-3.34} \times 10^{-4}$    \\ \hline
      CXB               & $\Gamma$\footnotemark[$\ddagger$]  & 1.4 (fixed)  \\
                        & $N$\footnotemark[$\S$]             & $8.14^{+0.31}_{-0.30} \times 10^{-4}$    \\ \hline  
                        & $\chi^2/d.o.f$                     & 120.96/110    \\ \hline 
    \\
   \multicolumn{2}{@{}l@{}}{\hbox to 0pt{\parbox{180mm}{\footnotesize
       \footnotemark[$*$] Temperature of the each component in keV.
       \par\noindent
       \footnotemark[$\dagger$] Normalization in the $apec$ model for each component scaled with a factor $1/400 \pi$. \\
                     $N=\frac{1}{400 \pi} \int n_{\rm e} n_{\rm H} dV / [ 4 \pi (1+z)^2 D_{\rm A}^2 ] \times 10^{-14}$ cm$^{-5}$ arcmin$^{-2}$,\\
                                where $D_{\rm A}$ is the angular diameter distance to the source.
       \par\noindent
       \footnotemark[$\ddagger$] Photon index of the power-law component.
       \par\noindent
       \footnotemark[$\S$] Normalization in the power-law component in photons keV$^{-1}$ cm$^{-2}$ s$^{-1}$ at 1 keV
     }\hss}}
    \end{tabular}
  \end{center}
\end{table*}

\subsection{Analysis of the temperature profile across the relic}
\label{sub:shock}
We investigate the temperature structure across the radio relic which is the candidate shock regions.
Assuming that a shock is located at the relic outer edge, we chose regions as shown in the top panel of figure \ref{fig:1} by green annular.
We search for point sources with Chandra data, whose spatial resolution is better than Suzaku.
In order to reduce the contamination and CXB systematic errors, we exclude $2'$ radius circular regions centered by the position of a point source (light blue circles in the figure \ref{fig:1} top panel) whose flux is more than $5 \times 10^{-15} {\rm \ erg/s/cm^{2}}$ (2.0-10 keV).
We fit the spectrum of each region by the following model;
\begin{eqnarray}
  constant*[apec_{\rm LHB} + phabs*(apec_{\rm MWH} &+& powerlaw_{\rm CXB} \nonumber \\
                                              &+& apec_{\rm ICM})],
\end{eqnarray}
where, $apec_{\rm ICM}$ represents the emission from the ICM.
We fixed all parameters of the background components ($apec_{\rm LHB}$, $apec_{\rm MWH}$ and $powerlaw_{\rm CXB}$) to the values derived from the background field analysis in subsection \ref{sub:back}.
We add $powerlaw_{\rm AGN}$ in case of the model fitting for the central region, because there is an active galactic nuclei (AGN) which can not be excluded.
Therefore, we fit the spectrum of the central region by the following model;
\begin{eqnarray}
  constant*[apec_{\rm LHB} &+& phabs*(apec_{\rm MWH}+powerlaw_{\rm CXB} \nonumber \\
                        &+& apec_{\rm ICM} +powerlaw_{\rm AGN})].
\end{eqnarray}
We assume $N_{\rm H} = 8.38 \times 10^{19} {\rm cm^{-2}}$ for the Galactic absorption \citep{Willingale13}.
The redshift and abundance of the ICM components are fixed to 0.0704 and 0.3 solar, respectively.
We introduced a parameter $constant$ to correct for slight differences in normalization among the XIS sensors.
It is known that there is uncertainty in the gain with negligible energy dependence in Suzaku XIS \citep{Koyama07,Yamaguchi08,Bamba08}.
The $constant$ is fixed to be unity for XIS1 and allowed to vary freely for XIS0 and XIS3.
For the spectral fitting of the ICM component, we used the energy band of 0.7 - 7.0 keV.
However, the energy band of 1.7 - 1.8 keV were ignored, because the response matrix around Si-K edge had residual uncertainties. 
Systematic errors of CXB are estimated in a way same as in \citet{Itahana15}.
We assume the upper cutoff flux of $S_c=5 \times 10^{-15} {\rm \ erg/s/cm^{2}}$. 
From the size of each region, the calculated CXB fluctuations at the 90 \% confidence level are shown in table \ref{tab:shock}.
In addition, it is reported that the reproducibility of the NXB was 4.9 \% at the 90 \% confidence level \citep{Tawa08}.
We estimated CXB and NXB systematic errors taking into account of these uncertainties.

\begin{figure*}
 \begin{center}
  \includegraphics[angle=270,width=7cm]{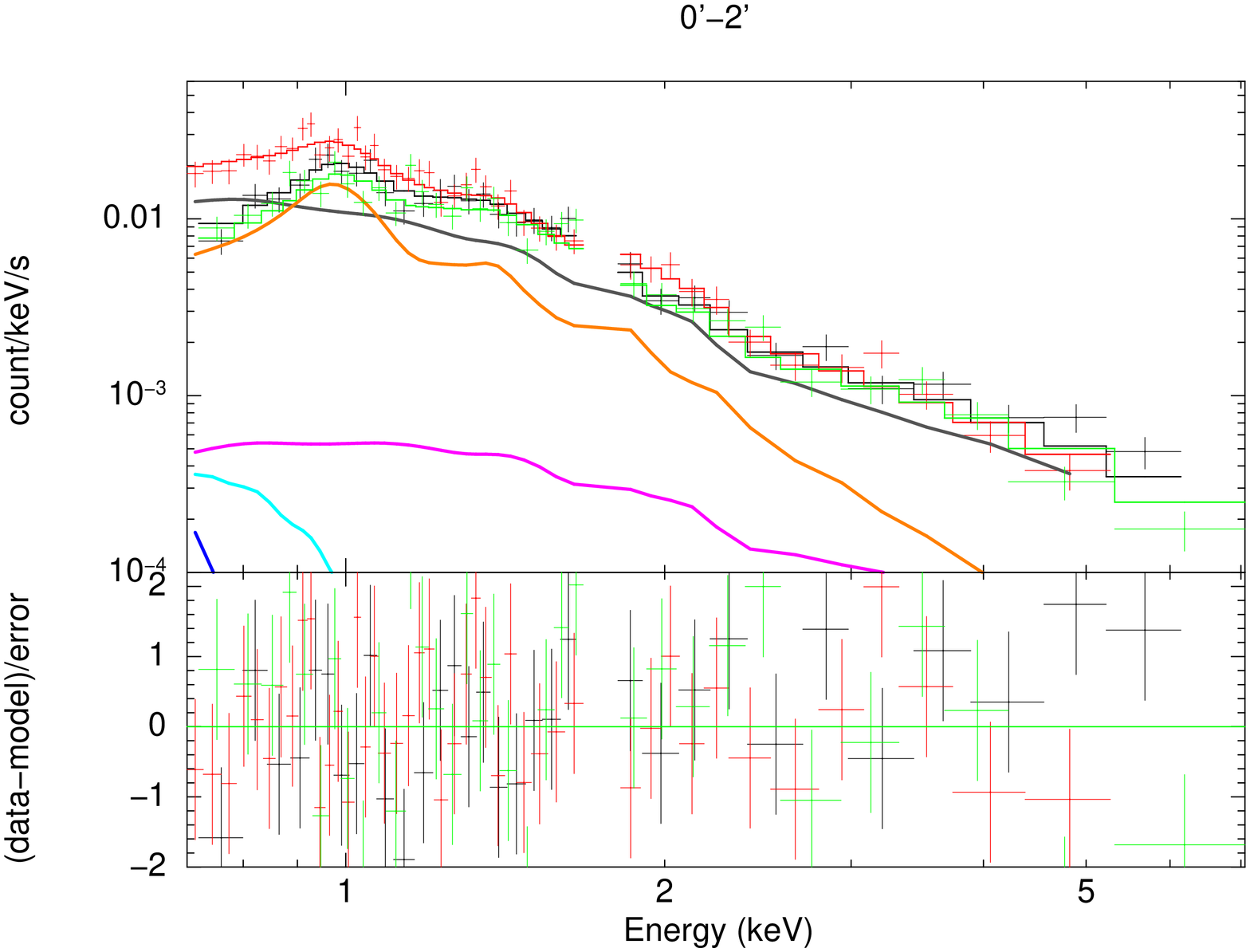}
  \includegraphics[angle=270,width=7cm]{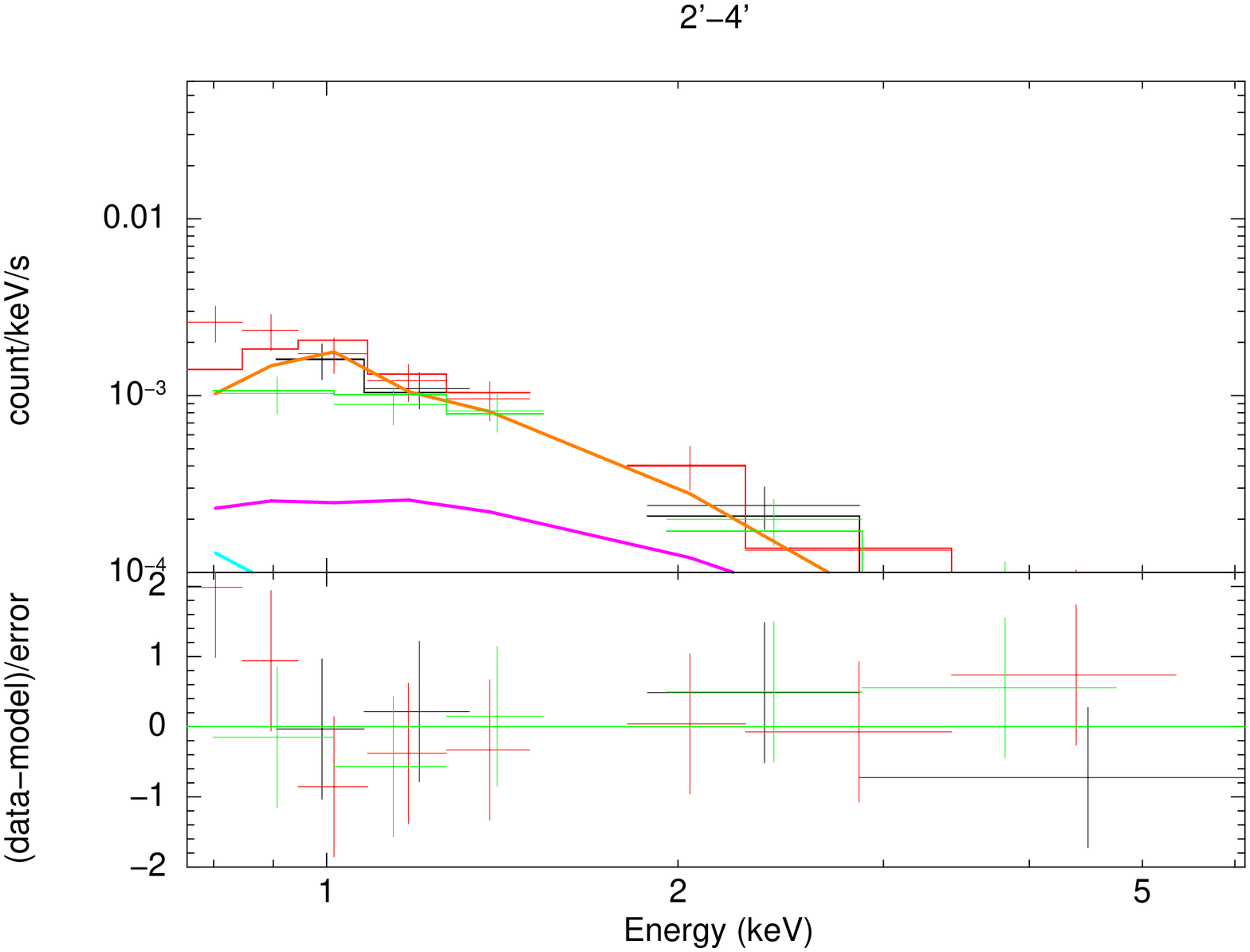}
  \includegraphics[angle=270,width=7cm]{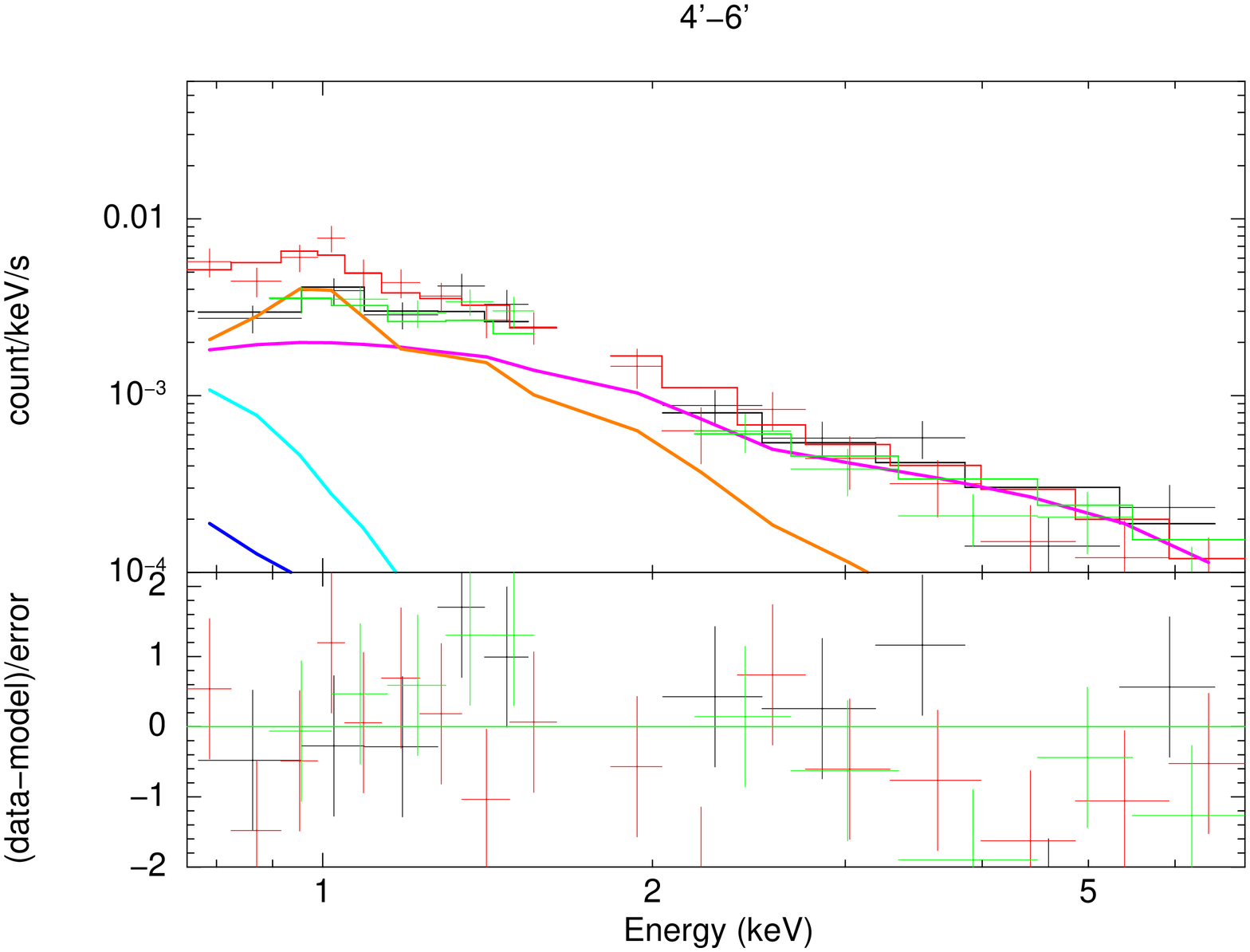} 
  \includegraphics[angle=270,width=7cm]{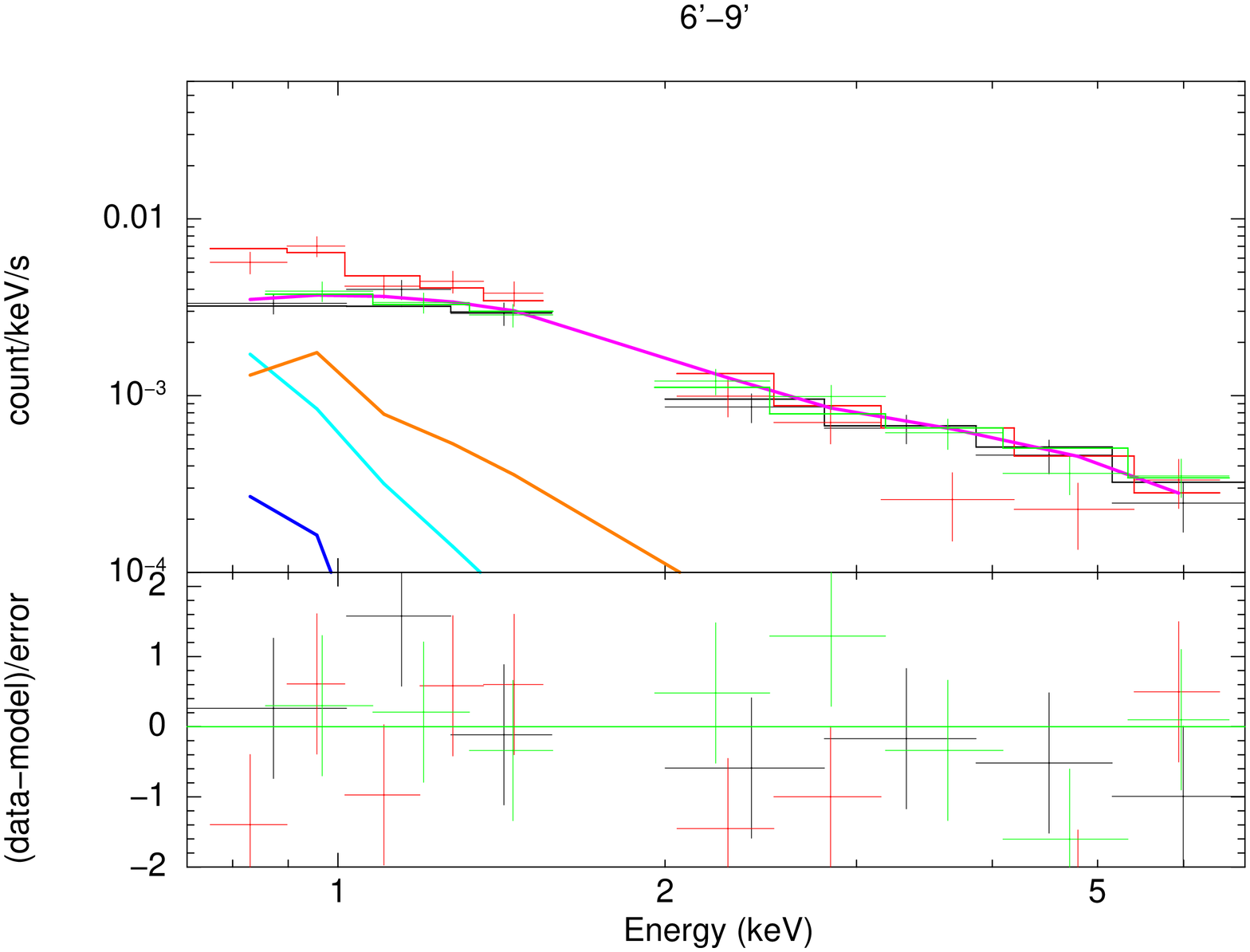}
  \includegraphics[angle=270,width=7cm]{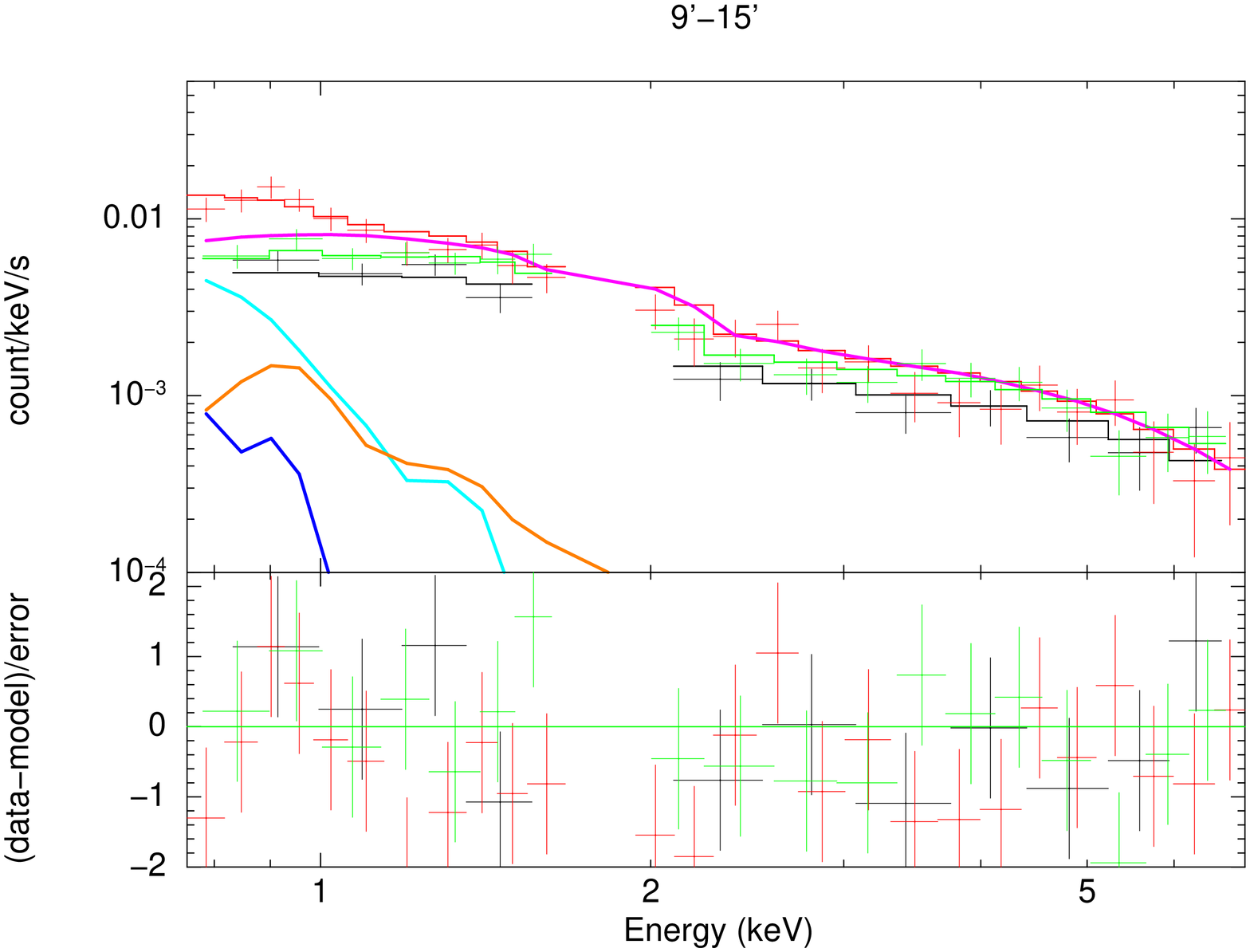}
 \end{center}
\caption{The XIS spectra of regions across the relic fitted with the model described in the text. The black, red and green crosses show the spectra of XIS0, XIS1,and XIS3, respectively. The blue, light blue, magenta, and orange solid histograms represent the LHB, MWH, CXB, and ICM components, respectively. The Gray solid line in the $0' -2'$ region spectra is an AGN component. }\label{fig:3}
\end{figure*}

\begin{figure}
 \begin{center}
  \includegraphics[angle=270,width=8cm]{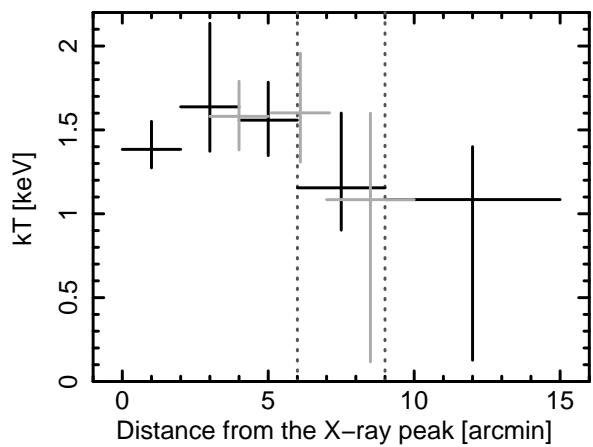} 
 \end{center}
\caption{The temperature profile across the radio relic. The horizontal axis represents the angular distance from the X-ray peak.  Black and light gray crosses shows the results from green and yellow regions of figure \ref{fig:1} top panel. Only statistical errors are displayed. The positions of the inner and outer edges of the relic are displayed by dark gray dotted lines. }\label{fig:4}
\end{figure}

Figure \ref{fig:3} shows the spectra of each region fitted with the above-mentioned model.
The ICM components of the outside region ($ > 6'$) are fainter than the CXB component. 
The resultant best-fit parameters are listed in table \ref{tab:shock}.
The first, second, and third errors are statistical, CXB systematic, and NXB systematic at the 90\% confidence level, respectively.
For the outside region of the radio relic ($9'-15'$), the ICM component is marginally detected if we consider only statistical errors.
Taking into account of systematic errors, however, the normalization of the $apec$ component is consistent with zero.
The temperature profile is shown in figure \ref{fig:4}, where the horizontal axis represents the angular distance from the X-ray peak.
In the central region, the best-fit parameters of $powerlaw_{\rm AGN}$ photon index and normalization are $\Gamma=2.19^{+0.11}_{-0.13} $ and $norm=1.66^{+0.32}_{-0.37} \times 10^{-2} {\rm \ photon/keV/s/cm^2}$, respectively, and flux of the AGN becomes $2 \times 10^{-13} {\rm \ erg/s/cm^2}$ (2.0-10 keV).
Here, we also measured the temperature in the central region with the Chandra data.
As a result, the obtained temperature becomes $kT=1.52^{+0.53}_{-0.23}$ keV.
This value is consistent with the Suzaku result taking account of statistical errors.
Though we did not find a significant temperature jump at the outer edge of the relic, the temperature decreases outward across the radio relic.

If the shock is not located just at the relic outer edge, the temperature in the pre- and post-shock regions could be over- and underestimated, respectively.
To check this, we measured the temperature of the yellow regions in figure \ref{fig:1} top panel, which are $1'$ shifted outward compared with green ones.
The obtained temperatures are shown with light gray crosses in figure \ref{fig:4}.
This suggests that the shock could be located $6'-9'$ from the X-ray peak and that the temperature in the $6'-9'$ region could be underestimated.

\begin{table*}
  \caption{Fitting results of region across the relic.}
  \begin{center}
    \begin{tabular}{lcccc} \hline
      Region   & $kT$ (keV)\footnotemark[$*$]        & normalization\footnotemark[$*$]         & $\chi^2/d.o.f$ & $\Delta_{\rm CXB}$ (\%)\footnotemark[$\dagger$] \\ 
       \hline
0-2'   & $1.38^{+0.17+0.04+0.01}_{-0.11-0.04-0.01}$ & $4.38^{+1.37+0.14+0.02}_{-1.21-0.18-0.02} \times 10^{-2}$  & 118.17/105     & 35 \% \\
2'-4'   & $1.64^{+0.50+0.41+0.01}_{-0.26-0.22-0.02}$ & $1.44^{+0.32+0.31+0.01}_{-0.31-0.27-0.01} \times 10^{-2}$ & 8.1/13     & 65 \% \\ 
4'-6'  & $1.56^{+0.23+0.16+0.04}_{-0.22-0.19-0.04}$ & $3.67^{+0.71+0.87+0.08}_{-0.70-0.75-0.09} \times 10^{-3}$ &  38.93/33      & 25 \% \\
6'-9' & $1.15^{+0.44+0.72+0.04}_{-0.25-0.72-0.05}$ & $5.83^{+3.94+4.20+0.03}_{-2.56-0.88-0.03} \times 10^{-4}$ &  36.02/21      & 19 \% \\ 
9'-15' & $1.08^{+0.32+0.00+0.00}_{-0.96-0.95-0.00}$ & $1.95^{+1.79+6.89+0.39}_{-1.40-1.53-0.12} \times 10^{-4}$ & 44.73/51       & 11 \% \\ \hline
3'-5' &   $1.58^{+0.21+0.20+0.02}_{-0.20-0.20-0.02}$ & $7.05^{+1.17+1.33+0.10}_{-1.15-1.22-0.11} \times 10^{-3}$ & 27.81/31      &  32 \%    \\
5'-7' &   $1.60^{+0.36+0.18+0.05}_{-0.29-0.23-0.06}$ & $2.17^{+0.63+0.73+0.08}_{-0.62-0.64-0.09} \times 10^{-3}$ &  38.93/31     &  24 \%     \\
7'-10' &  $1.08^{+0.51+0.00+0.00}_{-0.98-0.01-0.00}$ & $2.62^{+2.18+2.08+0.22}_{-2.04-2.39-0.19} \times 10^{-4}$ & 47.08/35      &  17 \%     \\ \hline

         \multicolumn{2}{@{}l@{}}{\hbox to 0pt{\parbox{180mm}{\footnotesize
       \footnotemark[$*$] The first, second, and third errors are statistical, CXB systematic, and NXB systematic, 
             respectively.
       \par \noindent
        \footnotemark[$\dagger$] CXB fluctuations at the 90 \% confidence level estimated.
     }\hss}}
    \end{tabular}
    \label{tab:shock}
  \end{center}
\end{table*}

\subsection{Analysis of the Surface Brightness Edge}
\label{sec:CF}
\begin{figure}
 \begin{center}
  \includegraphics[width=7.5cm]{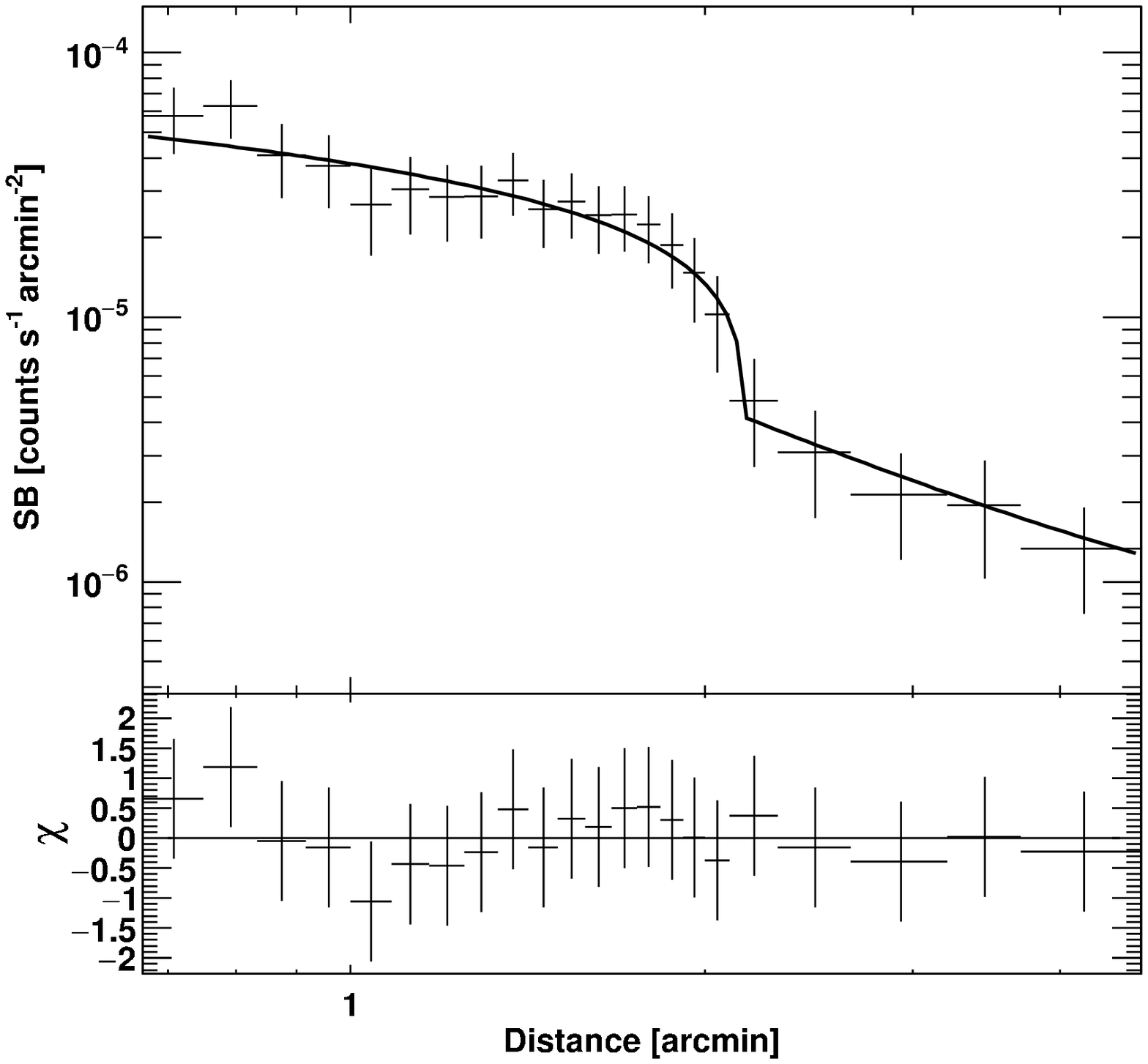}
 \end{center}
\caption{ The surface brightness profile across the surface brightness edge. The profile was binned to a minimum signal-to-noise ratio of 2 per bin. The best fit model described in the text is displayed with the black line.  }\label{fig:5}
\end{figure}

The Chandra image shows a surface brightness edge at the distance of $\sim 2'$ from the X-ray peak toward the west.
We investigate temperature and density structures of this region with Suzaku and Chandra, respectively.
First, we extract the surface brightness profile from Chandra data and estimate the density ratio across the surface brightness edge.
We exclude compact sources detected in the 0.5-7.0 keV band with the ``CIAO'' task ``wavdetect'' using scales of 1, 2, 4, 8, 16 pixels and cutting at the $3 \sigma$ level.
We used PROFFIT \citep{Eckert11} to extract and fit the profile.
The green sector shown in the bottom panel of figure \ref{fig:1} is used to extract the surface brightness profile.
The instrumental backgrounds are subtracted, and we used the energy band of 0.5 - 2.0 keV.
We determine the sky background component by fitting a $constant$ model for the outer region ($5.0 - 7.0$ arcmin) of the profile.
In the following analysis, the sky background component was fixed to this value.
We assume the following density model:
\begin{eqnarray}
n(r) = \left\{ \begin{array}{ll}
    n_1\left( \frac{r}{R_f} \right)^{-\alpha_1} , &  r<R_f \\
    n_1\frac{1}{C} \left( \frac{r}{R_f} \right)^{-\alpha_2} , & r>R_f
  \end{array}\right.
\end{eqnarray}
where $n(r)$ is the electron number density at the radius $r$, $n_i$ ($i=1, 2$) is the density, and $R_f$ is the radius of the location of the discontinuity in arcmin. $\alpha_i$ ($i=1, 2$) is powerlaw index, $C$ is the density contrast ($n_1/n_2$), and 1 and 2 denote the inside and outside region, respectively.
The profile and best-fit model ($\chi^2/d.o.f= 4.77/17 $) are shown in the figure \ref{fig:5}.
The fitting results are summarized in table \ref{tab:4}.
We obtain the density contrast of $C=2.44^{+2.50}_{-1.22}$ at the edge ($R_f=2.'14^{+0.16}_{-0.15}$).

\begin{table*}
  \caption{Fitting results of the surface brightness profile.}
  \begin{center}
    \begin{tabular}{ccccccc} \hline            
  $\alpha_1$ & $\alpha_2$ & $R_f$\footnotemark[$\S$] & $C$ & $S_0$\footnotemark[$*$] & $const$\footnotemark[$\dagger$] & $\chi^2/d.o.f$ \\ 
  \hline
$0.46^{+0.51}_{-0.61}$ & $1.66^{+1.15}_{-1.61}$ & $2.14^{+0.16}_{-0.15}$ & $2.44^{+2.50}_{-1.22}$ & $1.26^{+0.77}_{-0.58} \times 10^{-5}$ & $4.37 \times 10^{-7} $ (fixed) & $4.77/17  $ \\ 
  \hline  
\multicolumn{2}{@{}l@{}}{\hbox to 0pt{\parbox{180mm}{\footnotesize
\footnotemark[$\S$]The radius of the location of the discontinuity in arcmin.
\par\noindent
\footnotemark[$*$]Normalization of the surface brightness. 
\par\noindent
\footnotemark[$\dagger$]The sky background component.
\par\noindent

}\hss}}
    \end{tabular}
    \label{tab:4}
  \end{center}
\end{table*}

\begin{figure}
 \begin{center}
  \includegraphics[width=8cm]{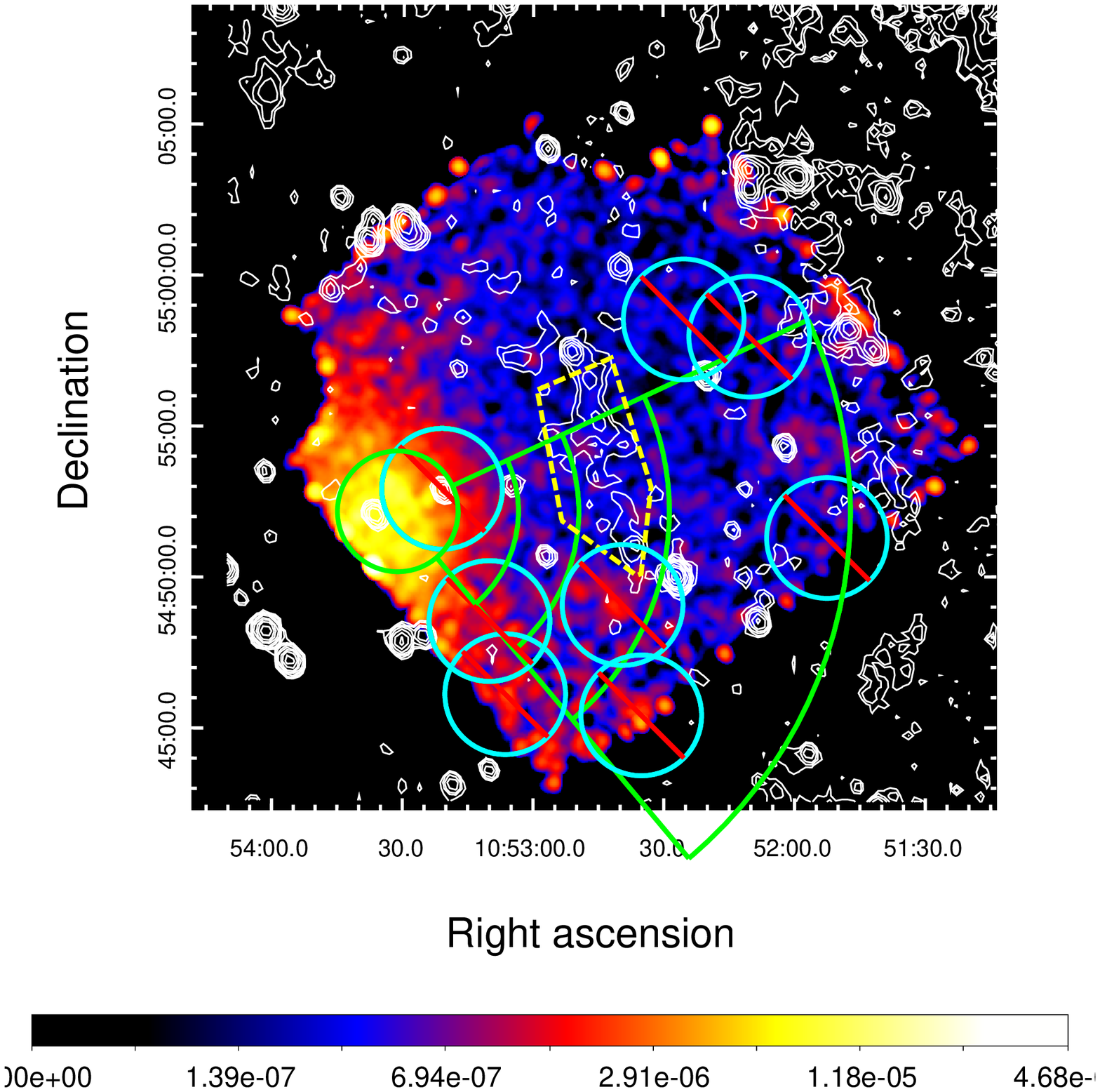} 
 \end{center}
\caption{The XIS image and radio contours same as in the top panel of figure \ref{fig:1}, but overlaid with green regions utilized in the surface brightness edge analysis. The radius of annulus region are $2'$, $4'$, $6'$, $9'$, and $15'$. Yellow region was used to search for inverse Compton component, whose size is $4.50 \times 10^{-3} {\rm deg^{2}}$. The light blue circles are excluded regions of a point source. }\label{fig:6}
\end{figure}

\begin{figure}
 \begin{center}
  \includegraphics[angle=270,width=8cm]{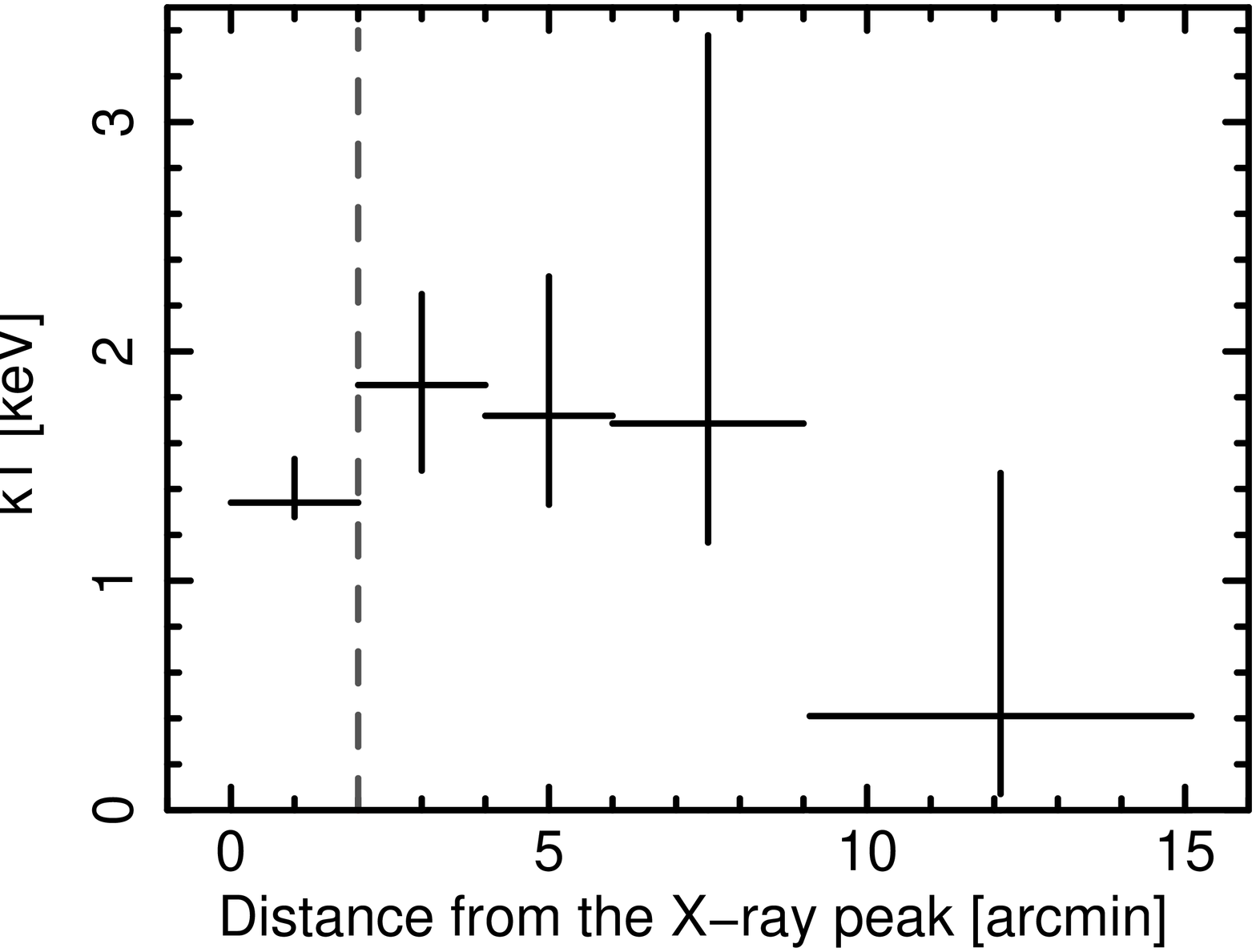} 
 \end{center}
\caption{The temperature profile across the surface brightness edge. The position of the surface brightness edge is displayed by dark gray dotted line.  }\label{fig:7}
\end{figure}

Secondly, we investigate the temperature structures around the surface brightness edge with Suzaku.
Regions used in this analysis are displayed with green in figure \ref{fig:6}.
The radius of annulus region are $2'$, $4'$, $6'$, $9'$, and $15'$, respectively.
We fit the spectrum of each region by the model same as in subsection \ref{sub:shock}.
The resultant best-fit parameters are listed in table \ref{tab:cold}, and the temperature profile is shown in figure \ref{fig:7}.
We obtained the temperature difference $T_1/T_2 = 0.72^{+0.24}_{-0.15}$ at the surface brightness edge, using the results of $0'-2'$ and $2'-4'$ regions.
This indicates that the temperature increases outward across the edge.

\begin{table*}
  \caption{Spectral fitting results of regions across the surface brightness edge.}
  \begin{center}
    \begin{tabular}{lcccc} \hline
      Region   & $kT$ (keV)\footnotemark[$*$]        & normalization\footnotemark[$*$]         & $\chi^2/d.o.f$ & $\Delta_{\rm CXB}$ (\%)\footnotemark[$\dagger$] \\ 
       \hline
0-2'   & $1.34^{+0.19+0.01+0.00}_{-0.06-0.01-0.00}$ & $4.39^{+0.96+0.11+0.00}_{-0.95-0.10-0.00} \times 10^{-2}$  & 113.15/100     & 39 \% \\
2'-4'   & $1.85^{+0.40+0.23+0.03}_{-0.37-0.55-0.03}$ & $1.66^{+0.26+0.21+0.01}_{-0.26-0.40-0.01} \times 10^{-2}$ & 17.24/12     & 53 \% \\ 
4'-6'  & $1.71^{+0.61+0.27+0.00}_{-0.39-0.28-0.00}$ & $2.72^{+0.86+1.07+0.09}_{-0.92-1.15-0.09} \times 10^{-3}$ &  18.47/11      & 34 \% \\
6'-9' & $1.72^{+1.88+1.51+0.21}_{-0.52-0.36-0.06}$ & $1.59^{+0.70+0.90+0.10}_{-0.68-0.70-0.07} \times 10^{-3}$ &  13.18/21      & 24 \% \\ 
9'-15' & $0.41^{+1.06+0.02+0.01}_{-0.34-0.23-0.01}$ & $8.75^{+4.48+2.89+0.60}_{-7.56-3.10-0.57} \times 10^{-4}$ & 18.25/24       & 15 \% \\ \hline

         \multicolumn{2}{@{}l@{}}{\hbox to 0pt{\parbox{180mm}{\footnotesize
       \footnotemark[$*$] The first, second, and third errors are statistical, CXB systematic, and NXB systematic, 
             respectively.
       \par \noindent
        \footnotemark[$\dagger$] CXB fluctuations at the 90 \% confidence level estimated.
     }\hss}}
    \end{tabular}
    \label{tab:cold}
  \end{center}
\end{table*}

\subsection{Search for the Non-thermal Inverse Compton X-rays from the Radio Relic}
\label{sec:IC}

The same electron population attributed to the radio relic are expected to emit non-thermal X-rays via the inverse Compton scattering of the CMB photons. 
We selected the yellow region in figure \ref{fig:6}, whose shape and size are similar to those for the radio flux measurement in \citet{vanWeeren11}.
First, we determine the thermal ICM temperature in this region by fitting the same spectral model as in subsection \ref{sub:shock}.
We obtain the ICM temperature of $kT=1.22$ keV.
In the following fit, we fix the temperature of the ICM component to this value because the inverse Compton component most-likely to be much weaker than the thermal ICM one.
Second, we fit the extracted spectra from the radio relic region by the following model;
\begin{eqnarray}
  constant*[apec_{\rm LHB} &+& phabs*(apec_{\rm MWH} +powerlaw_{\rm CXB} \nonumber \\
                        &+& apec_{\rm ICM}+powerlaw_{\rm IC})],
\end{eqnarray}
where, $powerlaw_{\rm IC}$ represents the inverse Compton component.
We fixed all parameters of the background components ($apec_{\rm LHB}$, $apec_{\rm MWH}$ and $powerlaw_{\rm CXB}$) to the values derived from the background field in subsection \ref{sub:back}.
We assume two cases for the $powerlaw_{\rm IC}$ photon index ($\Gamma$); 2.0 or 3.8.
Note that there is no observational information about the radio spectra of this relic.
Assuming that a simple diffusive shock acceleration (DSA) theory with the obtained Mach number ($M_{\rm X}$) from our results, the photon index becomes $\Gamma = \alpha +1 = 3.8$ (see sec.\ref{sec:Dshock}).
However, this implies spectra much steeper than typical radio relics.
On the other hand, a simple DSA dose not seem to hold for some relics.
For example, in toothbrush cluster, the obtained $M_{\rm X} \sim 1.5$ from Suzaku data \citep{Itahana15} is significantly lower than the value estimated from the radio data \citep{van Weeren2012}.
If a similar situation occurs, $\Gamma$ can be significantly smaller than $3.8$.
Therefore, we assume $\Gamma = 2.0$ as an extreme case. 
Systematic errors of CXB and NXB were taken into account in the same way as in the former analysis.
The fitting results are summarized in table \ref{tab:IC}.
Though we did not detect the inverse Compton component, we obtain an upper limit of the inverse Compton component.
The resultant upper limits on the flux are $F_{\rm IC} < 1.9 \times 10^{-14} {\rm \ erg/s/cm^2}$ and $ < 8.2 \times 10^{-11} {\rm \ erg/s/cm^2}$ for $\Gamma=2.0$ and $\Gamma=3.8$, respectively, in 0.3-10 keV considering both the statistical and systematic errors for $4.50 \times 10^{-3} {\rm \ deg^{2}}$ area.

\begin{table*}
\caption{Spectral fitting results of the spectrum model with inverse Compton components.}
\begin{center}
\begin{tabular}{ccc} \hline
      components         & $kT$ (keV) or $\Gamma$                & normalization\footnotemark[$\ddagger$]              \\ \hline
      $apec_{\rm ICM}$     & $1.22$ (fixed)\footnotemark[$*$]       & $8.36^{+3.95+1.42+0.42}_{-4.49-3.62-0.64} \times 10^{-4}$     \\
      $powerlaw_{\rm IC}$  & $2.0$ (fixed)\footnotemark[$\dagger$]  & $1.95^{+1.43 \times 10^{12}+2.17 \times 10^{12}+45.25}_{-0.81-1.95-1.95}      \times 10^{-16}$ ($ < 2.60 \times 10^{-4}$\footnotemark[$\S$])      \\ \hline 
 & $\chi^2/d.o.f$ & $28.23/19$                      \\ \hline

      $apec_{\rm ICM}$     & $1.22$ (fixed)\footnotemark[$*$]       & $8.37^{+3.94+3.11+0.44}_{-3.81-3.64-0.65} \times 10^{-4}$     \\
      $powerlaw_{\rm IC}$  & $3.8$ (fixed)\footnotemark[$\dagger$]  & $1.11^{8.22 \times 10^{19}+0.92+47.61}_{-1.11-0.44-0.21}  \times 10^{-20}   $ ($ < 8.22 \times 10^{-1}$\footnotemark[$\S$])      \\ \hline 
 & $\chi^2/d.o.f$ & $28.23/19$                      \\ \hline
 \\

\multicolumn{2}{@{}l@{}}{\hbox to 0pt{\parbox{180mm}{\footnotesize
\footnotemark[$*$]The value obtained from spectral fitting of the relic region in figure \ref{fig:6}.
\par\noindent
\footnotemark[$\dagger$] Assumed values for the photon index.
\par\noindent
\footnotemark[$\ddagger$] Normalizations in the $apec$ code and powerlaw component are written in the same way 
                           as in table \ref{tab:BGD}.\\
                           The errors are represented as in table \ref{tab:shock}.\\
\footnotemark[$\S$] Upper limits of normalization.
}\hss}}
\end{tabular}
\label{tab:IC}
\end{center}
\end{table*}

\section{Discussion}

\subsection{Temperature in the cluster central region}
 $L_{\rm X} - kT$ and $\sigma_{\rm v} - kT$  relations \citep{Wu98,Xeu00,Novicki02,Hilton12} are useful to investigate the physical status of galaxy clusters.
We compare the temperature of our results in the cluster central region with the expected one from $L_{\rm X} - kT$ and $\sigma_{\rm v} - kT$  relations.
First, \citet{Hilton12} reported a $L_{\rm X} - kT$ relation as follows,
\begin{eqnarray}
  \log \left( E^{-1}(z) \frac{L_{\rm X}}{ {\rm \ erg/s}}\right) &=& (44.67 \pm 0.09)  \nonumber \\
  &+& (3.04 \pm 0.16 ) \log \left( \frac{kT}{ 5 {\rm \ keV}} \right) \nonumber \\
  &-& ( 1.5 \pm 0.5 ) \log(1+z)   ,
\label{eq;kT}
\end{eqnarray}
where $L_{\rm X}$ is X-ray luminosity in the 0.1-2.4 keV band, $z$ is redshift and $E(z)= [\Omega_0 (1+z)^3 + \Lambda_0]^{1/2} $.
Using equation (\ref{eq;kT}), the deduced temperature of this cluster is $kT = 3.04 \pm 1.08 {\rm \ keV}$ ($1 \sigma$ confidence level) with $L_{\rm X[0.1 - 2.4 keV]}=0.96 \times 10^{44} {\rm \ erg/s}$ \citep{popesso04}.
We also checked the deduced temperature from another $L_{\rm X} - kT$ relation \citep{Novicki02}.
The obtained temperature is $kT=2.98 \pm 1.24 {\rm \ keV}$ and consistent with the above-mentioned one.
Next, \citet{Wilson16} reported a $\sigma_{\rm v} - kT$ relation as follows,
\begin{eqnarray}
\log \left( \frac{\sigma_{\rm v}}{\rm 1000 \ km/s} \right) &=& (0.02 \pm 0.05)  \nonumber \\
  &+& (0.86 \pm 0.14) \log \left(\frac{ kT}{\rm 5 \ keV} \right)  \nonumber \\
  &-& (0.37 \pm 0.33)\log E(z) ,
\label{eq:sigma-kT}
\end{eqnarray}
where $\sigma_{\rm v}$ is velocity dispersion.
Using equation (\ref{eq:sigma-kT}), the deduced temperature is $kT=2.98 \pm 1.10 {\rm \ keV}$  ($1 \sigma$ confidence level) with $\sigma_{\rm v} = 665^{+51}_{-45} {\rm \ km/s}$ \citep{Aguerri07}.
Again, we also checked the deduced temperature from another $ \sigma_{\rm v} - kT$ relation \citep{Wu98}.
The resultant temperature is $kT=3.28 \pm 0.28  {\rm \ keV}$ and consistent with the above-mentioned one.
The estimated temperatures from both relations are higher than our result at the center considering both statistical and systematic errors ($1.38^{+0.11}_{-0.07}$ keV ; $1 \sigma$ confidence level ).

Cool core components should be removed in the analysis with $L_{\rm X} -kT$ and $\sigma_{\rm v} - kT$ relations.
If this point is not appropriately treated, we will have lower temperature than expected.
In fact, we see a weak temperature decrease towards the center.
However, the cool core clusters usually have a centrally-peaked distribution of metal abundance.
We checked the abundance in the central region with the spectral fitting where the abundance is  free parameter.
As a result, the obtained value ($Z=0.13^{+0.14}_{-0.05} Z_{\odot}$) is not so high and we did not find such a feature in this cluster.
Thus, our result is not likely due to a cool core in this cluster.
As another possibility, it is likely that this cluster is not in dynamical equilibrium.
Numerical simulations of cluster mergers shows that the ICM temperature decreases due to an adiabatic expansion after the collision \citep{Ishizaka97,Takizawa99}.

The $\sigma_{\rm v} -kT$ relation is derived from most of galaxy clusters which are regarded as in dynamical equilibrium.
If the cluster is in an adiabatic expansion phase after the collision, the velocity dispersion could be high, and the ICM temperature decreases.
As a result, the expected temperature from the $\sigma_{\rm v} -kT$ relation could be higher than the obtained temperature from observation.
In fact, the measured temperature is lower than the expected one.
Thus, the above results also suggest that this cluster might be in an adiabatic expansion phase after the collision.

\subsection{Candidate Shock}
\label{sec:Dshock}
Though we did not find a significant temperature jump at the outer edge of the relic, figure \ref{fig:4} shows that the temperature decreases outward across the relic.
This suggests the existence of the shocks.
We derive the Mach number ($M_{\rm X}$) from the temperature difference of the shock candidate region using the a Rankine-Hugoniot relation as follows,
\begin{eqnarray}
  \frac{T_{\rm post}}{T_{\rm pre}} = \frac{5M_{\rm X} ^4 + 14M_{\rm X} ^2 -3}{16M_{\rm X} ^2},
\label{eq:R-H}
\end{eqnarray}
where $T_{\rm pre}$ and $T_{\rm post}$ are temperature of pre-shock and post-shock regions, respectively, and we assume that the specific heat ratio $\gamma \equiv 5/3$.
We did not find a significant temperature jump at the relic outer edge where the shock is expected.
From figure \ref{fig:4}, if shock exists, it could be located in $6'-9'$ from the X-ray peak.
Considering Suzaku's spatial resolution, the temperatures of the post-shock region could be underestimated.
Therefore, we used the temperatures of regions $4'-6'$ and $9'-15'$ in figure \ref{fig:1} top panel for pre- and post-shock, respectively.
As a result, we obtain the $M_{\rm X}=1.44^{+0.48+0.14+0.03}_{-0.91-1.34-0.04}$, where the first, second, and third errors are statistical, CXB systematic, and NXB systematic at the 90\% confidence level, respectively.
Some theoretical studies suggest that there are difficulties in particle acceleration at low Mach number shocks.
For example, \citet{Vink14} reported that shocks of the Mach number less than $\sqrt{5}$ cannot accelerate particles.
Our results seem to contradict this because the existence of radio relics is evidence of the accelerated particles.

Assuming a shock with the above-mentioned Mach number exists at the radio relic, we can calculate the expected radio spectral index.
We can obtain the shock compression ($C \equiv \rho_2/\rho_1$) from a Mach number.
If the electrons are accelerated by DSA, the energy spectrum of electrons becomes a power-law with index $p$ ($n(E)dE \propto E^{-p} dE$), which is related with shock compression and given by $p=(C+2)/(C-1)$.
For our result, the obtained shock compression is $C=1.6$ because the $M_{\rm X}=1.44$.
From this value, the power-law index becomes $p=5.7$.
This means that the spectral index of synchrotron radio at the injection region is $\alpha_{\rm inj}= (p-1)/2=2.3$.
The integrated radio spectrum is steeper by 0.5 compared to $\alpha_{\rm inj}$ \citep{Pacholczyk70,Miniati02}.
As a result, $\alpha= \alpha_{\rm inj} +0.5 =2.8$.  
Thus, we used the photon index of $\Gamma= \alpha +1 =3.8$ in subsection \ref{sec:IC}.

If the Mach number of the shock is estimated from future radio observations, we might be able to get crucial information about the particle acceleration process around the radio relic in this cluster.
Unfortunately, because no radio spectral information has been obtained so far, we cannot compare the Mach number of the shock from radio observations with our result.
Radio observations at other frequency bands are necessary for this cluster.

In the figure \ref{fig:7}, the temperature profile south of the relic suggests that a decrease of the temperature around $9'$ from the X-ray peak.
The obtained Mach number is $M_{\rm X}=3.27^{+3.41+1.80+0.25}_{-1.76-0.46-0.08}$ using the temperature difference of the regions $6'-9'$ and $9'-15'$.
This is higher than the aforementioned value around the relic, which suggests that Mach number varies along the radio relic, though errors are large.

\subsection{Surface Brightness Edge} 
We estimate the pressure profile across the surface brightness edge.
This is important to investigate the physical states of the edge.
In figure \ref{fig:7} and \ref{fig:5}, we see that the temperature and density increases and decreases outward across the surface brightness edge, respectively.
This suggests that this structure is not associated with a shock.
We calculate the pressure ratio ($P_1/P_2$) at the edge.
Using our results about the temperature ratio ($kT_1/kT_2=0.72^{+0.24}_{-0.15}$) and density ratio ($n_1/n_2=2.44^{+2.50}_{-1.22}$) at the edge from Suzaku and Chandra data, we obtain $P_1/P_2= 1.76^{+1.89}_{-0.95}$.

In the X-ray image, cold fronts are seen as an edge-like structure.
The temperature and the density are discontinuous but the pressure is continuous across a cold front.
Our results show that the temperature and the density show an increase and a sharp decrease outward across the surface brightness edge, respectively.
Additionally, the pressure could be continuous across the edge.
These suggest that the edge could be a cold front.
If the existence of a cold front is true, this could be evidence that the galaxy cluster RXC J1053.7+5453 experienced merger with other cluster (or group).
Considering the overall morphology of X-ray emission in the Chandra image, the location of a possible cold front, and the orientation and location of the relic, it seems that this cluster is in an east-west merger event.
Alternatively, the thermal pressure ratio at the edge could be really more than unity.
It cannot be ruled out that the thermal pressure is discontinuous at the edge.
In this case, if the pressure across the surface brightness edge is in equilibrium, other forms of pressure sources such as cosmic-rays are necessary.

\subsection{Magnetic Field Strength in the Radio Relic}
\label{sec:mag}

We constrain the magnetic field strength in the relic.
\citet{Blumenthal70} derived following equations for the synchrotron and inverse Compton emissions from an electron population with a power-law energy spectrum at frequency $\nu_{\rm Synch}$ and $\nu_{\rm IC}$, respectively.
\begin{eqnarray}
\frac{dW_{\rm Synch}}{d\nu_{\rm Synch}dt} = \frac{4 \pi N_0 e^3 B^{(p+1)/2}}{m_e c^2}\left(\frac{3e}{4 \pi m_e c} \right)^{(p-1)/2}a(p) \nu_{\rm Synch}^{-(p-1)/2} , \nonumber \\
\\
\frac{dW_{\rm IC}}{d\nu_{\rm IC}dt} = \frac{8 \pi^2 r_0^2}{c^2} h^{-(p+3)/2} N_0 (kT_{\rm CMB})^{(p+5)/2} F(p) \nu_{\rm IC}^{-(p-1)/2}, \nonumber \\ 
\label{eq:fd}
\end{eqnarray}
where $N_0$ is the normalization, $p$ is the power-law index of the electron spectrum ($N( \gamma) = N_0 \gamma^{-p}$ ;$\gamma$ is the Lorentz factor of the electron), $r_0$ is the classical electron radius, $h$ is the Planck constant, and $T_{\rm CMB}$ is CMB temperature ($T_{\rm CMB}=2.73(1+z)$). 
The function $a(p)$ and $F(p)$ are given as follows \citep{Blumenthal70}:
\begin{eqnarray}
a(p)= \frac{2^{(p-1)/2} \sqrt{3} \Gamma \left(\frac{3p-1}{12} \right)  \Gamma \left(\frac{3p+19}{12} \right)  \Gamma \left(\frac{p+5}{4} \right)}{8 \pi^{1/2} (p+1)  \Gamma \left(\frac{p+7}{4} \right)} , \\ 
F(p) = \frac{2^{p+3}(p^2+4p+11)  \Gamma \left(\frac{p+5}{2} \right)  \zeta \left(\frac{p+5}{2} \right) }{(p+3)^2(p+5)(p+1)} . 
\label{eq:aF}
\end{eqnarray}
The magnetic field strength in the radio relic can be estimated from the comparison of the observed flux density of the synchrotron and inverse Compton emissions by the relation $S_{\rm Synch}/S_{\rm IC}=\frac{dW_{\rm Synch}}{d\nu_{\rm Synch}dt}/\frac{dW_{\rm IC}}{d\nu_{\rm IC}dt}$ \citep{Ferrari08,Ota08,Ota14,Akamatsu16}.

First, in case of $\Gamma=2.0$, we derive the upper limit of the inverse Compton flux density as $S_{\rm IC}<2.22 \times 10^{-10} {\rm Jy} $ at 10 keV ($\nu_{\rm IC} = 2.4 \times 10^{18} $ Hz) from the spectral analysis for the non-thermal power-law component ($\Gamma=2.0$ in sec.\ref{sec:IC}).
Comparing this limit with the radio flux density of the relic as $S_{\rm Synch}=15$ mJy at 1382 MHz \citep{vanWeeren11}, the magnetic field strength becomes $B > 0.73 {\rm \mu G}$.
This value is similar with those of other relics ($B > 0.1 - 1 {\rm \mu G}$).

Secondly, when $\Gamma=3.8$, we obtain $B > 2.00 {\rm \ \mu G}$ using $S_{\rm IC} < 1.11 \times 10^{-8} {\rm Jy}$ as well as case of $\Gamma=2.0$.
This lower limit of magnetic field strength is rather high.
Additionally, if this is true, the energy density of the magnetic field could be higher than the thermal one.
We will discuss this problem in the next subsection.

\subsection{Energy Densities in the Radio Relic Region}
\label{sec:EneDens}
From our results, we estimate the energy densities of the thermal ICM, non-thermal electrons, and magnetic field.
These are fundamental and important parameters for radio relic studies.
We estimate in the same way as in \citet{Itahana15}.
For simplicity, we assume that the radio relic region is a cylinder whose radius and height are 284 kpc and 243 kpc, respectively.
The estimated electron number density of the thermal ICM in the radio relic region becomes $n_e = 3.12^{+0.78}_{-1.08} \times 10^{-5} {\rm \ cm^{-3}}$ from the normalization of the $apec_{\rm ICM}$ model (table \ref{tab:IC}).
The energy density of the thermal ICM is $U_{\rm th}= 3n_e kT/2 \mu =1.52^{+1.10}_{-0.45} \times 10^{-13} {\rm \ erg/cm^3}$ with the obtained $n_e$ and the temperature of this region ($kT=1.22^{+0.80}_{-0.33}$ keV), assuming that the mean molecular weight is $\mu= 0.6$.

Next, we estimate the energy densities of the magnetic field ($U_{\rm mag}= B^2/8 \pi$) and the non-thermal electrons ($U_{\rm e}$) for each photon index ($\Gamma=$2.0 , 3.8).
In case of $\Gamma=2.0$, the energy density of the magnetic field is $U_{\rm mag} > 2.1 \times 10^{-14} {\rm \ erg/cm^3}$.
As a result, $U_{{\rm mag}}/U_{\rm th} > 0.14 $.
This means that the magnetic energy could be more than a ten percent of the thermal one.
The energy densities of the non-thermal electrons corresponding to 0.3-10 keV X-ray band (or, $5.6 \times 10^2 < \gamma < 3.3 \times 10^3$) is $U_{e } < 7.8 \times 10^{-16} {\rm erg/cm^3}$.
Therefore, $U_{{\rm e}}/U_{\rm th} < 5.1 \times 10^{-3} $ for $\Gamma=2.0$, 
although we did not included the contribution from lower energy electrons in these calculation, which could be dominant in the energy density of the non-thermal electron populations.

On the other hands, when $\Gamma=3.8$, we obtain $U_{\rm mag} > 1.6 \times 10^{-13} {\rm \ erg/cm^3}$ and $U_{e }< 5.6 \times 10^{-12} {\rm erg/cm^3}$.
As a result, $U_{{\rm mag}}/U_{\rm th} > 1.00 $ and $U_{{\rm e}}/U_{\rm th}< 36.7$.
This means that the magnetic field energy density might be higher than the thermal one.
This seems to be quite odd and makes us suspicious of the underlying assumptions.
For example, DSA does not hold and hence $\Gamma \neq 3.8$, or, the electron spectrum is significantly deviated from a single power-law form.
Another possibility is that the Mach number derived from our results is seriously underestimated owing to projection effects and/or limited spatial resolution.
In this case, the actual electron spectrum should be flatter.

\section{Conclusions}
We observed the field around the radio relic in the galaxy cluster RXC J1053.7+5453 ($z=0.0704$) with Suzaku and Chandra.
From the Suzaku XIS analysis, we measured temperature of this cluster for the first time and the resultant temperature around the center is lower than the values expected from both its X-ray luminosity and velocity dispersion.
Additionally, we found the temperature decrease outward across the relic and derived the Mach number ($M_{\rm X} \sim 1.4$) assuming that it is due to a shock associated with the radio relic.
Because no radio spectral information has been obtained, we cannot compare the Mach number of the shock derived from radio observations with our result.
We found a surface brightness edge at the distance of $\sim 2'$ from the X-ray peak toward the west in the Chandra X-ray image.
We performed X-ray spectral and surface brightness analyses around the edge with the Suzaku and Chandra data, respectively.
The obtained surface brightness and temperature profiles suggest that the edge is not a shock but a cold front, because the pressure ratio at the edge is consistent with unity.
However, it cannot be ruled out that the thermal pressure is discontinuous across the edge.
In this case, to balance the pressure across the surface brightness edge, other forms of pressure source, such as cosmic-rays, are necessary. 
We searched for the non-thermal inverse Compton component in the relic region.
Though we did not detect the inverse Compton component, we obtained the upper limit on the flux are $ 1.9 \times 10^{-14} {\rm \ erg/s/cm^2}$ and $  8.2 \times 10^{-11} {\rm \ erg/s/cm^2}$ for $\Gamma=2.0$ and $\Gamma=3.8$, respectively.
The lower limit of magnetic field strength becomes $ 0.7 {\rm \ \mu G}$ and $ 2.0 {\rm \ \mu G}$ for $\Gamma=2.0$ and $\Gamma=3.8$, respectively.
In case of $\Gamma=3.8$, however, odd physical situation occurs, which seems to be unlikely.

\begin{ack}
The authors would like to thank T. Akahori, S. Shibata, and H. Ohno for helpful comments. 
We are also grateful to the Suzaku operation team for their support in planning and executing this observation.
MT is supported in part by Japan Society for the Promotion of Science (JSPS) KAKENHI Grant Number 26400218. 
HA and FZ acknowledge the support of NWO via a Veni grant. SRON is supported financially by NWO, the Netherlands Organization for Scientific Research.
Support for this work was provided by the National Aeronautics and Space Administration through Chandra Award Number GO5-16140X issued by the Chandra X-ray Observatory Center, which is operated by the Smithsonian Astrophysical Observatory for and on behalf of the National Aeronautics Space Administration under contract NAS8-03060. R.J.W. is supported by a Clay Fellowship awarded by the Harvard-Smithsonian Center for Astrophysics.
\end{ack}


\end{document}